# Three-dimensional imaging of biological cells using surface plasmon coupled emission

**Anik Mazumder** ,[a,b] **Mohammad Mozammal** ,[a]
**and Muhammad Anisuzzaman Talukder** [a,]*

[a]Bangladesh University of Engineering and Technology,
Department of Electrical and Electronic Engineering, Dhaka, Bangladesh
[b]United International University, Department of Computer Science and Engineering,
Dhaka, Bangladesh

## Abstract

**Significance:** Biological cell imaging has become one of the most crucial research interests because of its applications in biomedical and microbiology studies. However, three-dimensional (3D) imaging of biological cells is critically challenging and often involves prohibitively expensive and complex equipment. Therefore, a low-cost imaging technique with a simpler optical arrangement is immensely needed.

**Aim:** The proposed approach will provide an accurate cell image at a low cost without needing any microscope or extensive processing of the collected data, often used in conventional imaging techniques.

**Approach:** We propose that patterns of surface plasmon coupled emission (SPCE) features from a fluorescently labeled biological cell can be used to image the cell. An imaging methodology has been developed and theoretically demonstrated to create 3D images of cells from the detected SPCE patterns. The 3D images created from the different SPCE properties at the far-field closely match the actual cell structures.

**Results:** The developed technique has been applied to different regular and irregular cell shapes. In each case, the calculated root-mean-square error (RMSE) of the created images from the cell structures remains within a few percentages. Our work recreates the base of a circular-shaped cell with an RMSE of $\lesssim 1.4\%$. In addition, the images of irregular-shaped cell bases have an RMSE of $\lesssim 2.8\%$. Finally, we obtained a 3D image with an RMSE of $\lesssim 6.5\%$ for a random cellular structure.

**Conclusions:** Despite being in its initial stage of development, the proposed technique shows promising results considering its simplicity and the nominal cost it would require.

**Keywords:** cell imaging; three-dimensional imaging; surface plasmon resonance; surface plasmon coupled emission.

## 1 Introduction

Human cells appear in many shapes and sizes. The cell shapes change according to their specialized functions, such as the mechanical imbalance between the forces exerted on them by the external environment and the intracellular components.[1] For example, intracellular osmotic pressure, polymerization of actin networks, and cytokinesis lead to the gradual metamorphosis of cells during embryogenesis.[1] In addition, during cell migration, in response to chemical or mechanical signals, cells show amoeboid mobility, which involves morphological expansion and retraction.[2] For example, lymphocytes change their shape to squeeze past tightly packed tissue cells to reach the site of infection.[3] Neutrophils change their shapes to swallow bacteria and viruses.[3] Many other biological functions require the deformation of cells.

*Address all correspondence to Muhammad Anisuzzaman Talukder, anis@eee.buet.ac.bd

Furthermore, different types of cells look different, but cells of the same kind look similar, maintaining a uniform shape. Notably, dead or cancer cells are misshaped, and they appear as a collection of cells that significantly vary in shape and size.[4] Therefore, cell shape and size can be essential parameters for the diagnosis and prognosis of many diseases like cancer. At present, many people around the world are dying because of cancer. It has been identified as the leading cause of death worldwide.[5] Therefore, an inexpensive imaging technique is essential for detecting abnormal cells. Furthermore, cell imaging has become a crying need during the COVID-19 pandemic for its potential application in virology to diagnose the virus and study its impact on human cells.[6]

Currently, several biological cell imaging techniques exist, such as scanning electron microscopy (SEM),[7–9] Raman spectroscopy,[10–12] and total internal reflection fluorescence microscopy (TIRFM).[13,14] Although SEM-based techniques usually provide a much greater resolution than many other imaging techniques, they often destroy live cells because of the high-energy electron beam used.[15] SEM is expensive, requires extensive equipment, and must be housed in an electric, magnetic, and vibrational interference-free area. Special training is necessary to prepare samples for SEM and operate it. In contrast, Raman spectroscopy is a label-free, nondestructive, and noninvasive technique. However, its sensitivity is low, and a weak Raman signal leads to a long acquisition time.[16] In TIRFM, the depth of the optical section is typically $\lesssim$100 nm,[13] which is much smaller than the height of most cells. The fluorescence resonance energy transfer (FRET) technique is famous for single-molecule or cell height ($h$) detection.[17] However, FRET is limited to applications for only $h \lesssim 10$ nm.

Recently, surface plasmon coupled emission (SPCE)-based techniques have been used to detect biological samples and analyze biomolecular interactions.[18–20] SPCE-based techniques probe the near-field interaction between fluorophores and surface plasmons at the metal surface. In such techniques, a biological sample labeled with fluorophores is placed on a metal film deposited on a glass prism. When the fluorophores are excited, highly directional light is emitted from the sample–metal layer interface.[21,22] The emission angle ($\theta_r$) depends on the fluorophore position on the sample and the thickness and refractive index of the sample layer. If the thickness of the sample layer varies, $\theta_r$, intensity, and the number of rings ($N_r$) also change.[23] Additionally, the SPCE far-field patterns change when the fluorophore position varies within the sample.[24,25] Therefore, the highly directional emission at certain angles and the emission intensity contain essential information about the sample environment and the change of state of the system.[26] Without needing any microscope, a camera can detect the emission cone, and the sample can be determined by observing different emission features. SPCE-based techniques are comparatively inexpensive, light, and have minimal hardware requirements.

This work proposes a technique for creating three-dimensional (3D) images of biological cells by observing different SPCE features from the fluorescently labeled cell samples. This work also demonstrates the application of the proposed technique using a finite-difference time-domain (FDTD) numerical technique for varying shapes and sizes of cells. For 3D imaging, we determine the ($x$, $y$, $z$) coordinates of the fluorophores lying on the cell surface. SPCE initiated by the fluorophores is converged on an image plane using converging lenses. The center of a converged spot on the image plane corresponds to a fluorophore position on the $xy$ plane. It is found that $s$-polarized excitation shows better accuracy than the $p$-polarized one. The cell's height can first be estimated within a range using the information about $N_r$, $\theta_r$, and rings' relative dominance in the far-field. To further narrow down the estimated $h$, the full-width at half-maximum (FWHM) of the spots' electric field intensity distribution is used. The ratio of the FWHM from $s$-polarized excitation to that from $p$-polarized excitation ($R_c$) shows an approximately constant behavior for an $h$ irrespective of the cell shape and size. Therefore, $h$ and $z$ coordinates are calculated using the predetermined $R_c$ variation with $h$ and $z$ coordinates.

This work recreates the base of a circular-shaped cell with a root-mean-square error (RMSE) $\lesssim$1.4%. In addition, the images of irregular-shaped cell bases have an RMSE of $\lesssim$2.8% only. The proposed methodology shows a precision comparable to traditional fluorophore locating techniques with significantly less complexity and cost. One such method is 3D fluorophore localization based on gradient fitting,[27] which offers fluorophore localization RMSE of <10 nm in the lateral directions and <40 nm in the axial direction. In this work, we simulated different cell structures and found RMSE of <20 nm in both lateral directions and the axial direction. Furthermore, traditional fitting algorithms require a complex iterative process and do not consider lateral directions while determining $h$.[27,28] However, the proposed method avoids such computational complexity and considers lateral dimensions while detecting $h$. Additionally, conventional cell microscopy techniques are limited by the numerical aperture (NA) of the microscopes, which is not the case for this work.

## 2 Proposed Imaging Technique

The proposed imaging technique requires the implementation of the experimental setup given in Fig. 1. First, the cell is labeled by immersing in a solution of fluorophores. The labeled cell is then placed on the SPCE slide. A laser source illuminates the cell from the sample side in the reverse Kretschmann (RK) configuration. The excitation process is similar to that presented in Ref. 29. The laser source's linearly polarized light becomes circularly or elliptically polarized after passing through a quarter-wave plate. The circularly polarized light passes through a polarization cube (PC1), reflecting the vertical but transmitting the horizontal component. Thus $p$-polarized (vertical) and $s$-polarized (horizontal) excitation lights are separated to excite the sample one at a time. Two shutters, S1 and S2, control the sequential excitation. Shutter speeds can be controlled using microcontrollers. A second polarization cube (PC2) and mirrors ensure the normal incidence of $s$- and $p$-polarized light on the cell. Upon sequential $s$- and $p$-polarized illumination, the emissions from the fluorophores excite surface plasmons in the metal layer, which eventually result in highly directional conical emissions through a hemispherical prism.[30]

A 50-nm-thick silver layer is used as the plasmon containing metal layer. Silver is chosen as the metal layer for its enhanced capability to support surface plasmons in the visible wavelength region compared to other noble metals.[31,32] Also the imaginary part of the silver's dielectric function is the smallest compared to other noble metals that support surface plasmons,[33] resulting in less dissipation of surface plasmon polariton waves.[34] Therefore, silver stands out as a suitable metal for SPCE. Here the silver layer has a 50-nm thickness as the coupling of the fluorescence energy into the metal is optimum at this thickness.[31] A 10-nm glass is used as a spacer between the sample and metal layer to reduce the fluorophores' nonradiative quenching.[35] A hemispherical prism ensures normal incidence of the emitted light on the prism-air interface, providing zero deviation of the emitted light. As $h$ is usually >100 nm, RK configuration is necessary to directly excite the fluorophores on the cell surface by the incident light from the sample side. On the other hand, in Kretschmann–Raether (KR) configuration, fluorophores only within 100 nm from the metal surface are excited.[26] In addition, the light should be incident at a particular angle from the prism side in the KR configuration, which requires additional control equipment, making it complex to implement.

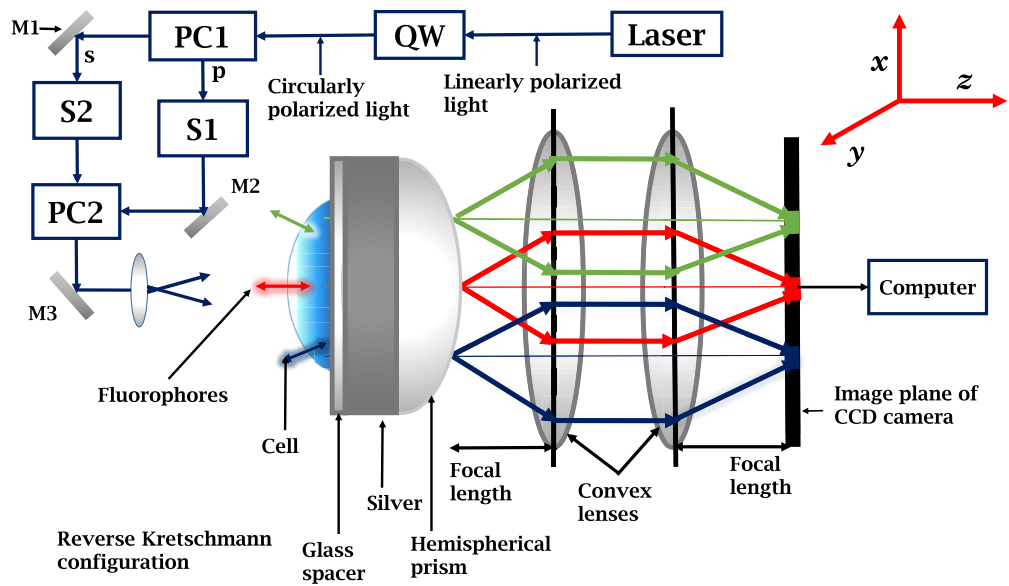

**Fig. 1** Schematic illustration of the proposed experimental setup.

The divergent emission converges into a circular spot through two convex lenses: $C_1$ and $C_2$. The converged light pattern is captured using a charge-coupled device camera and processed on a computer. From the circular spots' positions on the image plane, we can detect fluorophores' positions on the $xy$ plane. By observing the far-field pattern, we estimate $h$. The intensity and width of the circular spots on the image plane are used to precisely determine the fluorophores' positions in the $z$ direction.

In practice, the selection of fluorophore labels will be critical. Fluorophores are chemical substances that emit light upon excitation. They are excited at a specific frequency and emit light at a different frequency. Fluorophores' absorption and emission spectra depend on their concentration and the host medium.[36] Since an excited fluorophore emits a wavelength much longer than the physical dimension of the emitter itself, it can be mathematically considered equivalent to an electric dipole.[35] Therefore, in FDTD simulations, we have used an electric dipole source to replicate the excited fluorophore behavior. In recent years, rhodamine B and diI molecules have been frequently used for fluorescence imaging.[37] Both absorb light at ~550 nm and emit at ~570 nm.[38,39] Therefore, the laser must emit at ~550 nm to excite the fluorophore. The luminescence quantum yield of rhodamine B is ~0.5 to 0.68 in ethanol,[40] which is higher than other fluorophores in that medium. Ethanol has a refractive index close to that of the cell environment. In addition, rhodamine B is more photostable and pH insensitive.[41] All these make rhodamine B a better choice as a fluorophore. DiI molecules can also be used as they show similar behavior to rhodamine B.[39]

## 3 Simulation Setup

The basic setup of the SPCE structure, except for the sample layer, remains the same as the setup in Ref. 30. Rhodamine B fluorophore is modeled as an electric dipole in FDTD simulations. The operating wavelength of the dipole is set to 565 nm, the maximum emission wavelength of the fluorophore rhodamine B.[42] The refractive index values of all materials used in the sensor are considered at this wavelength. The refractive index data of silver and prism glass at the operating wavelength are taken from Ref. 43. The biological cell is modeled using the approach described in Ref. 37. A hemispherical or ovoid shape of 1 to 2 $\mu$m is used as a cell, surrounded by a membrane of 15 to 20 nm.[44,45] Therefore, a cell is a three-layer structure, where protoplasm is in the middle with the cell membrane on both sides. When the cell is placed on the spacer, the portion of the cell adjacent to the interface gets flattened. Therefore, it is assumed to be hemispherical. The overall refractive index of the cell is assumed as 1.38.[46,47] In real life, the intracellular components and the membrane may have a slightly different index. However, as their dimensions are much smaller than the light wavelength, a single index for the cell will not affect the results.[37] As a single index is considered for the cell structure, the membrane is included within the cell thickness. Besides considering hemisphere shapes for cells, ellipsoidal and irregular shapes have also been considered to determine the effectiveness of the proposed imaging technique.

The setup orientations in the Lumerical FDTD simulation environment are presented in Fig. 2. Material layers of the setup are shown in Fig. 2(c). Figure 2 shows that the emission is in the $z$ direction. 3D FDTD simulation is significantly time- and memory-consuming. Therefore, symmetric and antisymmetric boundary conditions are applied to minimize the simulation time and memory requirements whenever the structure appears symmetric for the fluorophore orientation. For fluorophores oriented in the $z$ direction, the electric field is tangent on the $xz$ and $yz$ planes, allowing symmetric boundary conditions in the negative $x$ and negative $y$ directions. In contrast, for fluorophores oriented in the $x$ direction, the electric field lies on the $xz$ plane but is normal to the $yz$ plane. Therefore, symmetric and antisymmetric boundary conditions are used in the negative $y$ and negative $x$ directions, respectively. The perfectly matched layer boundary condition is used in the $x$, $y$, and $z$ directions. Due to the applied boundary conditions, the FDTD simulation region appears semi-infinite in the $z$ direction and infinite in the $x$ and $y$ directions. The near-field profiles are recorded in the prism material on the $xy$ plane at 300 nm from the metal-prism surface. The simulation region is 6 $\mu m \times 6$ $\mu m \times 4$ $\mu m$. FDTD simulations are continued until the energy within the region decays to $10^{-5}$ of its initial value.

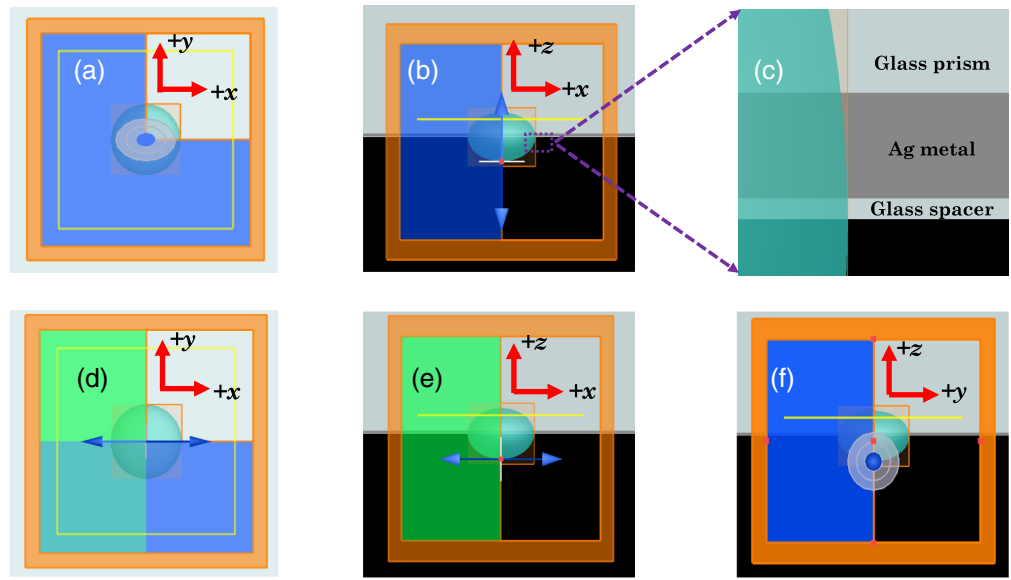

**Fig. 2** Simulation setup in the (a) $xy$ plane and (b) $xz$ plane with vertical dipole moment excited by $p$-polarized excitation; structure orientation in the $yz$ plane is the same as (b) and therefore not shown. (c) Materials in the $z$ direction. The material layer thicknesses are given in Sec. 2. Simulation setup in the (d) $xy$ plane, (e) $xz$ plane, and (f) $yz$ plane with horizontal dipole moment excited by $s$-polarized excitation.

FDTD simulation is a time-intensive numerical technique. Using a larger mesh size reduces computational time, resulting in less accuracy. Conversely, using a smaller mesh provides higher accuracy at the cost of a prolonged simulation time and memory. Here we have used smaller mesh sizes in the region of interest to resolve minor features, whereas larger mesh sizes in other areas. This nonuniform meshing helps us to complete FDTD simulations within a reasonable time without sacrificing the desired accuracy. We used a maximum grid size of 10 nm in the $x$ and $y$ directions and 5 nm in the $z$ direction in a region that encloses the cell and interfaces. The FDTD mesh settings in the regions around interfaces are overridden by closely spaced points to resolve the smallest features. A maximum grid size of 50 nm is used in other parts of the simulation region. The near-field emission profiles for a cell with a width or base diameter ($w$) of 2000 nm in the $x$ and $y$ directions and $h$ of 500 nm are shown in Fig. 3. The results comply with the findings presented in Ref. 37, validating our simulation setup and helping to comprehend the SPCE effectively in different directions.

## 4 Imaging Methodology

### 4.1 Determining Fluorophore ($x, y$) Coordinates and Recreating the Cell Base

In the proposed experimental setup, two convex lenses are used. The first one collects the highly directional SPCE emission and decomposes it into parallel plane waves. The second lens converges the plane waves onto a circular spot. Numerically, we calculate the near-field data and decompose it into plane waves to simulate the first lens. The plane waves propagate at different angles. Any plane waves with angles outside the lens's NA are then discarded. We assume NA = 1. Then using the inverse chirped $z$-transformation (ICZT), the remaining plane waves are recombined to create an image on an image plane like the second lens. ICZT converts the $k$ space representation to the spatial domain. It is like the inverse discrete Fourier transform (IDFT) but has some advantages over the IDFT. ICZT offers an arbitrarily fine representation of the field

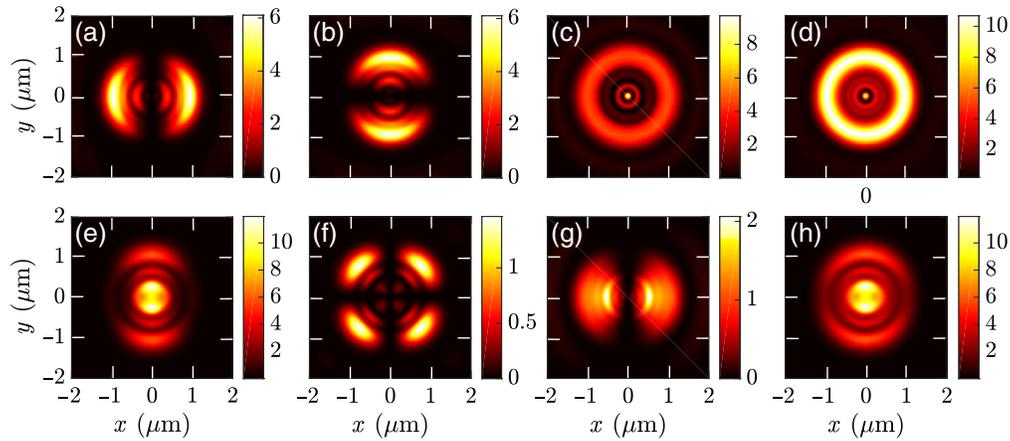

**Fig. 3** (a)–(d) Near-field profiles for vertical dipole moment in the $z$ direction. (e)–(h) Near-field profiles for horizontal dipole moment in the $x$ direction. (a), (e) $|E_x|^2$; (b), (f) $|E_y|^2$; (c), (g) $|E_z|^2$; and (d), (h) $|E_x|^2 + |E_y|^2 + |E_z|^2$.

profile in the space domain, which cannot be obtained using IDFT.[48] ICZT offers an accurate synthetic focusing of the fluorophore fields.[48]

By postprocessing the converged image plane data, the $(x, y)$ coordinates of the circular spot are calculated that indicate a fluorophore's position on the $xy$ plane. This process is carried out for both $s$- and $p$-polarized illuminations. It is found that the $s$-polarized excitation produces better accuracy in determining the coordinates.

### 4.1.1 Case study for locating fluorophore position on the xy plane

The approach described in Sec. 4.1 is applied to a cell with $w$ of 1400 nm in both the $x$ and $y$ directions and a $h$ of 600 nm in the $z$ direction. The fluorophore is placed at $(400, 300, -350)$ nm position. The setup is sequentially excited by $s$- and $p$-polarized light, simulated using horizontal and vertical dipoles, respectively. The obtained near-field and image-plane field profiles are shown in Fig. 4.

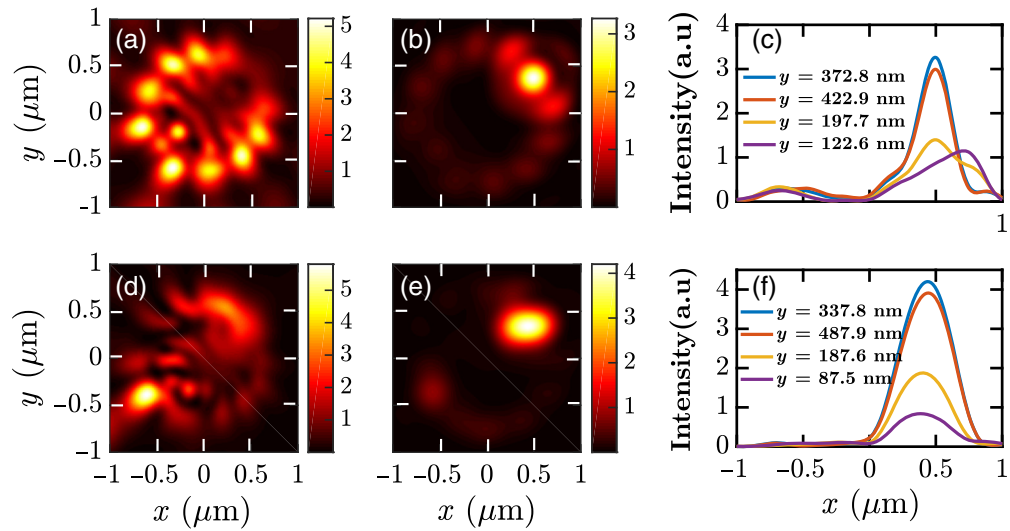

**Fig. 4** Emission intensity $|E_x|^2 + |E_y|^2 + |E_z|^2$ from a fluorophore located on a cell for (a)–(c) $p$-polarized excitation and (d)–(f) $s$-polarized excitation. Near-field profiles on the $xy$ plane for (a) $p$-polarized excitation and (d) $s$-polarized excitation. (b), (e) Field profiles after convergence of the SPCE emissions. (c), (f) Intensity profiles in the $x$ direction for different sections of $y$ on the image plane. In both cases, the $(x, y)$ coordinates of the maximum intensity indicate the dipole position on the $xy$ plane.

We note that it is challenging to determine the dipole position from the near-field image. However, we obtain an image with a circular or elliptical spot after decomposition and reconstruction. The maximum intensity of the image lies at the dipole position on the $xy$ plane. Therefore, we can find the maximum intensity by numerically processing the image plane data. Thus we can obtain the fluorophore position on the $xy$ plane. The calculated coordinates are (495.4, 372.8) nm and (437.9, 337.8) nm for the $p$- and $s$-polarized excitations, respectively, showing better accuracy for the $s$-polarized excitation.

### 4.1.2 *Accuracy in cell base imaging*

Practically, the cell will be labeled with numerous fluorophores, and each fluorophore position will indicate a point on the cell surface. Therefore, there will be multiple circular spots on the image plane. From these multiple coordinates, we can recreate the cell base. We simulated hemispherical, ellipsoidal, and irregular cell shapes and recreated the base shape in each case. First, we detected the center fluorophore, i.e., the fluorophore at the top. The variation of the spot width, i.e., the spot diameter ($d$), with the fluorophore position is shown in Fig. 5. We simulated a hemispherical-shaped cell with $w = 2000$ nm and $h = 400$ nm. The simulated ellipsoidal-shaped cell has a base diameter of 1200 nm in the $x$ direction ($w_x$), 1600 nm in the $y$ direction ($w_y$), and $h = 650$ nm. The polar angle around the base is denoted as $\phi$. Figures 5(a) and 5(b) show that the fluorophore at the center has a greater $d$ than that of other fluorophores in both the $x$ and $y$ directions, i.e., at $\phi = 0$ deg and $\phi = 90$ deg, respectively. We also found that the center fluorophore has a greater $d$ than other fluorophores at $\phi = 45$ deg. Therefore, the center fluorophore has the greatest $d$ among the fluorophores, allowing for detecting the center fluorophore.

After detecting the center fluorophore, fluorophore positions at the edges of the cell surface are determined. For this, the numerically determined fluorophores' coordinates are first sorted and then recorded in a matrix to locate the cell's base shape from the obtained matrix. The coordinates with the same $y$ but different $x$ values are placed in a row. Similarly, coordinates with the same $x$ but different $y$ values are placed in a column to construct the matrix. Figure 6 shows the workflow in detecting fluorophores. Using the workflow, we can now recreate the portion of the cell's surface that is in contact with the SPCE structure.

Table 1 shows the cell's different $(x, y)$ points for $s$- and $p$-polarized illuminations. The recreated base shape is shown in Fig. 7. Only a few points are shown in Table 1. To recreate the complete base, many other fluorophore positions have been detected. For $s$-polarized illumination, we find RMSE < 1.4% and <1.36% for $x$ and $y$ coordinates, respectively. For $p$-polarized excitation, the RMSE is <3.26% and <3.22% for $x$ and $y$ coordinates, respectively.

The $s$-polarized irradiation excites fluorophores oriented horizontal to the plane of the structure, whereas $p$-polarized excitation generates vertical dipole moment of the fluorophores. The $s$-polarized excitation creates both $s$- and $p$-polarized SPCE emissions, whereas $p$-polarized excitation creates only the $p$-polarized SPCE.[30,49] Better accuracy can be attributed to the horizontal dipole moment excited by $s$-polarized irradiation that satisfies the orthogonal alignment condition between the excitation polarization direction and the imaging axis.[50] Figure 8 depicts the fluorophore emission profiles on the $xz$ plane, showing significant radiation along the imaging axis for the horizontal dipole moment. However, there is negligible radiation for the vertical dipole moment in that direction. Fluorophore coordinates are determined from the image plane's maximum electric field intensity position. At the fluorophore position on the image plane, the electric field intensity is greater for the horizontal dipole moment excited by $s$-polarized illumination than the vertical dipole moment excited by $p$-polarized illumination,[50–52] as observed in Figs. 4(c), 4(f), and 8. In addition, the maximum intensity for the vertical dipole moment excited by $p$-polarized excitation is away from the fluorophore's position, leading to greater error, evident from the presented data in Table 1. Therefore, the $s$-polarized irradiation is more suitable for locating fluorophores.

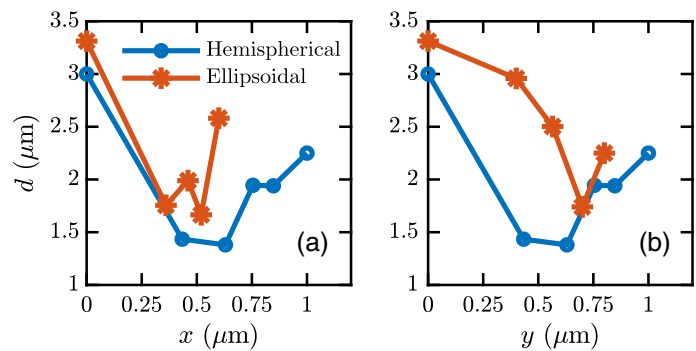

**Fig. 5** Converged intensity spot width, i.e., diameter ($d$), against fluorophore positions in the (a) $x$ direction ($\phi = 0$ deg) and (b) $y$ direction ($\phi = 90$ deg).

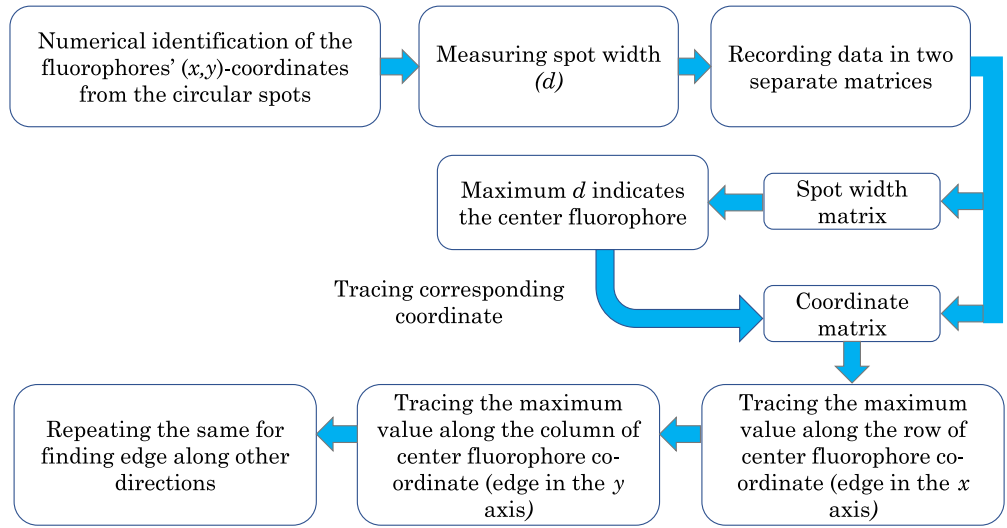

**Fig. 6** Schematic diagram of the cell base imaging process.

**Table 1** Fluorophore positions on the $xy$ plane for the simulated hemispherical cell.

| Actual $(x, y)$ (nm) | s-polarized excitation | | p-polarized excitation | |
|---|---|---|---|---|
| | Detected $(x, y)$ (nm) | Error $(x, y)$ (%) | Detected $(x, y)$ (nm) | Error $(x, y)$ (%) |
| (1000, 0) | (983.5, 0) | (1.65, 0) | (1026, 0) | (2.6, 0) |
| (965.5, 258) | (990.95, 260.26) | (2.64, 0.88) | (998.46, 267.78) | (3.41, 3.8) |
| (864, 500) | (875.83, 503) | (1.37, 0.6) | (890.85, 518.02) | (3.1, 13.6) |
| (705, 705) | (708, 708) | (0.43, 0.43) | (730, 730) | (3.55, 3.55) |
| (500, 864) | (503, 875.83) | (0.6, 1.37) | (518.02, 890.85) | (3.6, 3.1) |

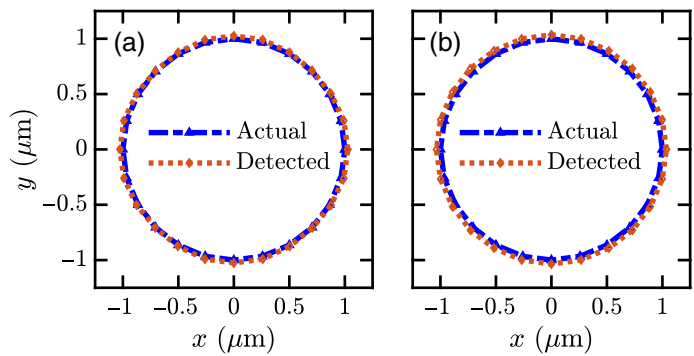

**Fig. 7** Actual and detected base of a hemispherical cell for (a) s-polarized and (b) p-polarized excitations.

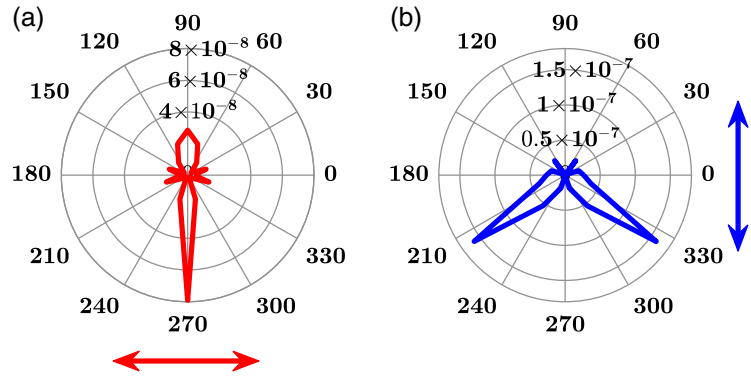

**Fig. 8** Emission patterns on the $xz$ plane of (a) horizontal dipole moment excited by $s$-polarized irradiation and (b) vertical dipole moment excited by $p$-polarized irradiation.

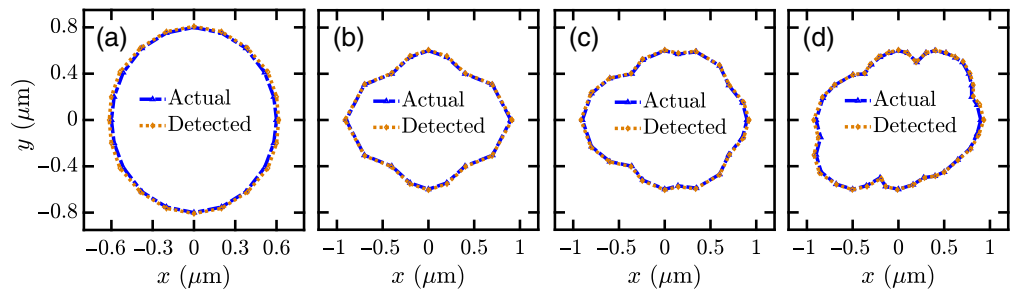

**Fig. 9** Images of (a) an elliptical base and (b)–(d) bases of some irregular cell shapes. All image points are obtained using $s$-polarized irradiation.

Figure 9(a) shows the base images of an ellipsoid cell shape. The ellipsoid cell has $w_x = 1200$ nm, $w_y = 1600$ nm, and $h = 650$ nm. The RMSE from $s$-polarized illumination is <2.78% and <1% for $x$ and $y$ coordinates, respectively. The RMSE from $p$-polarized illumination is <6.06% and <6.12% for $x$ and $y$ coordinates, respectively. We have also determined base images for irregular-shaped cells, as shown in Figs. 9(b)–9(d). The maximum RMSE for the recreated irregular cell bases is <2.8% for both the $x$ and $y$ coordinates. The images of irregular-shaped cells show the feasibility of the proposed technique for realistic scenarios, where cells, especially the unhealthy ones, may have irregular shapes and sizes.

### 4.2 Determining Fluorophore's $z$-Coordinates

In Sec. 4.1.2, we discussed how to determine the $(x, y)$ coordinates of the center fluorophore, which lies at the top of the cell. As it lies at the top of the cell, we can determine the cell's height from its $z$ coordinate. From the far-field SPCE patterns, $N_r$, $\theta_r$, and intensity are calculated. As fluorophores with horizontal dipole moment do not couple much energy to SPCE,[25] fluorophores with perpendicular dipole moment are chosen for SPCE feature extraction. Therefore, the $p$-polarized illumination is used for determining $h$. The far-field profiles are calculated by projecting the near-field profiles onto a 1-m-radius hemisphere. Using SPCE features in the far-field, we first determine the range within which $h$ lies. However, as these features are not the same for a fixed $h$ with different base sizes, they cannot determine $h$ precisely. Therefore, the relations between features obtained from different polarization configurations are also used to determine the precise $h$.

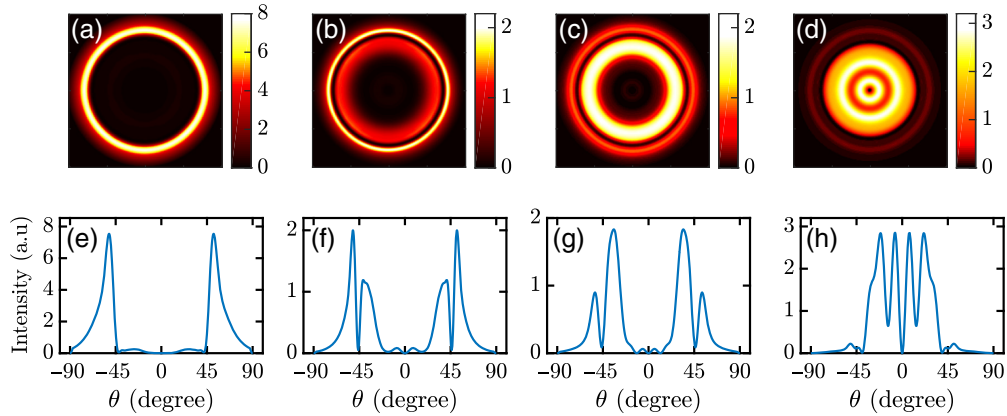

**Fig. 10** (a)–(d) Far-field SPCE profiles for different $h$. (e)–(h) Far-field intensity ($|E_x|^2 + |E_y|^2 + |E_z|^2$) profile versus angle. In all cases, $p$-polarized illumination is used. The cell height is (a), (e) 100 nm, (b), (f) 250 nm, (c), (g) 400 nm, and (d), (h) 800 nm.

The center fluorophore's intensity distribution on the image plane follows a Gaussian curve in the $x$ and $y$ directions. Therefore, a Gaussian fit to the intensity distribution is determined. Next, FWHM of the Gaussian distribution for the $s$- and $p$-polarized illuminations are calculated in the $x$ and $y$ directions. We note that the ratio of the FWHM from $s$-polarized illumination to that from $p$-polarized illumination ($R_c$) does not vary much for a particular $h$ irrespective of the base size and shape. Therefore, $h$ can be determined from $R_c$. The $z$ coordinates of the fluorophores, located at arbitrary positions on the cell membrane, are also determined using a similar approach.

### 4.2.1 Determining cell height

Figure 10 shows the far-field SPCE variations against the cell size for a fluorophore at the center of the cell. There is only one distinct far-field ring ($N_r = 1$) when $h = 100$ nm. However, when $h$ increases, $N_r$ increases as well. Here the surface plasmon resonances (SPRs) are coupled with the resonances in the cell sample layer, resulting in a Fabry–Pérot interferometer, i.e., two mirrors with a dielectric medium in between. The silver layer acts as one mirror, and the other is created due to internal reflections from the cell membrane–air interface.[25] When the cell is thicker than a critical value, SPCE is dominated by Fabry–Pérot modes, and below that, SPCE is dominated by SPR modes.[53] Therefore, multiple rings occur in the SPCE pattern above a critical thickness.[23,54] Hence, $N_r$ can determine $h$ within a range. The range of $h$ and corresponding $N_r$ are given in Table 2.

**Table 2** Ranges of $h$ and corresponding far-field SPCE features.

| Range of $h$ (nm) | $N_r$ | Dominant ring | Angle of the middle ring |
|---|---|---|---|
| <200 | 1 | Outer | — |
| 200 to 250 | 2 | Outer | ~38 deg to 40 deg |
| 250 to 350 | 3 | Middle | ~38 deg to 40 deg |
| 350 to 500 | 3 | Middle | ~30 deg to 35 deg |
| 500 to 750 | 3 | Outer < inner < middle | ~30 deg to 35 deg |
| 750 to 800 | 3 | Outer < inner < middle | ~28 deg to 30 deg |
| 800 to 1000 | 3 | Inner | ~28 deg to 30 deg |

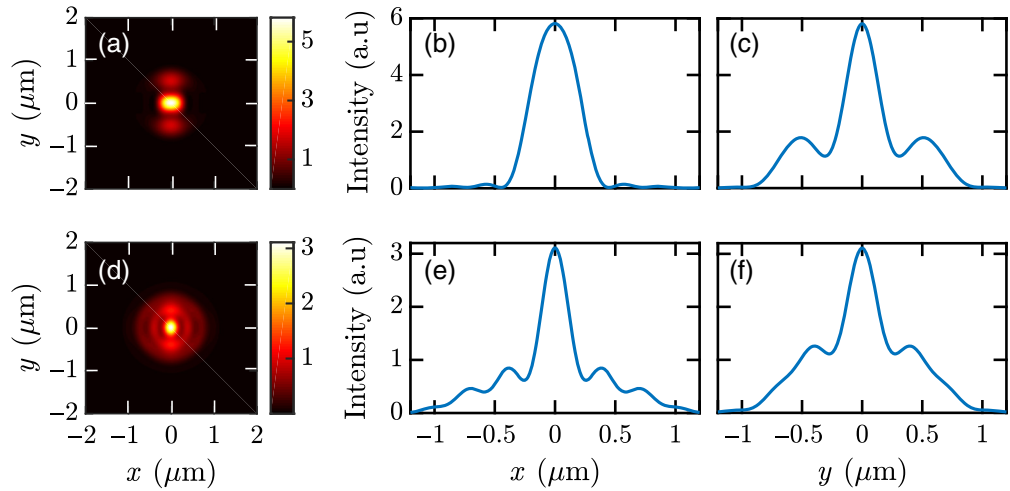

**Fig. 11** Spot created on the image plane from the emission of the fluorophore at the top of the cell surface for (a)–(c) *s*-polarized and (d)–(f) *p*-polarized excitations. (a), (d) Reconstructed spot on the image plane. Intensity distributions in the (b), (e) *x* direction and (c), (f) *y* direction.

Figure 10 also shows that the relative strength of rings changes with *h*. Figure 10(f) shows that the outer ring is stronger than the inner one and the innermost ring is almost invisible. In Fig. 10(g), the middle ring becomes dominant, whereas the innermost one is stronger than the outer ones in Fig. 10(h). Table 2 shows how the dominance of rings changes with *h*. Using the relative intensities of rings, we can further narrow down *h*. In Table 2, cells of different sizes are simulated with *h* up to 1000 nm. The angles of the outermost and the innermost rings are ~52 deg and ~7 deg, and they do not vary much with *h*. However, the angle of the middle ring changes with *h*. Using this information, *h* can be determined more precisely. The relations between features obtained from different polarization configurations can also determine *h* more precisely. For this purpose, a cell with $w_x = 1700$ nm, $w_y = 1300$ nm, and $h = 500$ nm is simulated. Figure 11 shows results with converged emission.

Figures 11 and 12 show that Gaussian curves fit the intensity distributions well. MATLAB curve fitting tool has been used to find the best fit to the intensity distributions using

$$f(x) = a \exp\left[-\frac{(x-b)^2}{c^2}\right].$$ (1)

Here *a* is the amplitude of the Gaussian distribution, *b* is the mean of the distribution, and *c* is the standard deviation of the Gaussian curve proportional to the FWHM of the intensity distribution. Figure 12 shows that the squared intensity (SI) distribution fits the Gaussian curve better. Therefore, the SI is used to determine the *c* parameters for the intensity distributions. The *c* parameters obtained from curve fitting are given in Tables 3 and 4. Here $c_{xs}$ and $c_{xp}$ are the *c* parameters from the intensity distribution fits in the *x* direction for *s*- and *p*-polarized illuminations, respectively. Furthermore, $c_{ys}$ and $c_{yp}$ are the *c* parameters in the *y* direction for $s$- and $p$-polarized illuminations. FWHM$_{xs}$ and FWHM$_{xp}$ are FWHMs of the intensity distributions in the $x$ direction for $s$- and $p$-polarized illuminations. FWHM$_{ys}$ and FWHM$_{yp}$ are FWHMs of intensity distributions in the $y$ direction for $s$- and $p$-polarized illuminations.

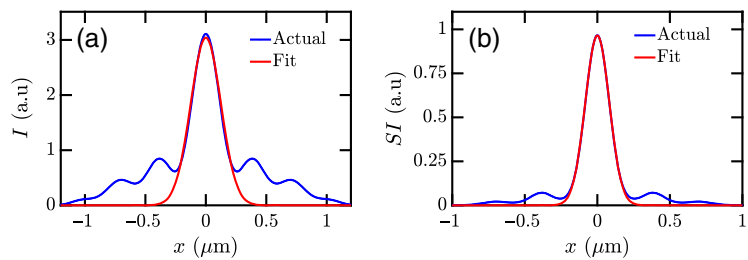

**Fig. 12** Gaussian fitting of the (a) intensity $I = (|E_x|^2 + |E_y|^2 + |E_z|^2)$ distribution and (b) SI $= (|E_x|^2 + |E_y|^2 + |E_z|^2)^2$ distribution.

**Table 3** Parameters obtained from $s$- and $p$-polarized illumination on a cell with $w_x = 1700$ nm, $w_y = 1300$ nm, and $h = 500$ nm.

| $c_{xs}$ (nm) | $c_{ys}$ (nm) | FWHM$_{xs}$ (nm) | FWHM$_{ys}$ (nm) |
|---|---|---|---|
| $c_{xp}$ (nm) | $c_{yp}$ (nm) | FWHM$_{xp}$ (nm) | FWHM$_{yp}$ (nm) |
| 191 | 139.3 | 450.4 | 325.4 |
| 123 | 150 | 285.2 | 350.4 |
| Ratio: 1.55 | | Ratio: 1.58 | |

**Table 4** Parameters obtained from $s$- and $p$-polarized illumination on a cell with $w_x = 1500$ nm, $w_y = 2000$ nm, and $h = 500$ nm.

| $c_{xs}$ (nm) | $c_{ys}$ (nm) | FWHM$_{xs}$ (nm) | FWHM$_{ys}$ (nm) |
|---|---|---|---|
| $c_{yp}$ (nm) | $c_{xp}$ (nm) | FWHM$_{yp}$ (nm) | FWHM$_{xp}$ (nm) |
| 186.8 | 132 | 445.4 | 310.4 |
| 120.4 | 124 | 280.2 | 290.2 |
| Ratio : 1.55 | | Ratio: 1.59 | |

In the previous sections, we showed how to determine the edges of a cell's base. Therefore, following the steps mentioned in the previous sections, we can determine $w$ in different directions. First, if $w_x > w_y$, the ratio of $c_{xs}$ and $c_{xp}$ is used to determine $h$, as shown in Table 3. Second, if $w_x < w_y$, the ratio of $c_{xs}$ and $c_{yp}$ is used to determine $h$, as shown in Table 4. In both cases, the ratio of $c$ parameters ($R_c$) is similar to the ratio of FWHM values. Therefore, $R_c$ can be interpreted as the ratio of FWHMs of intensity distributions for $s$- and $p$-polarized excitations. We also note that $R_c$ is almost the same for different cells with the same $h$. Therefore, $h$ can be determined from $R_c$ irrespective of the cell shape and size. In this work, different cell sizes have been considered to calculate $R_c$ with equal $h$.

Table 5 shows a few other examples for different $w_x$, $w_y$, and $h$. We find that $R_c$ does not vary much if $h$ remains constant irrespective of the cell size or shape. However, $R_c$ varies when $h$ changes. For a particular $h$, $R_c$ values have been calculated for different $w$. Then the average of $R_c$ values have been calculated for a particular $h$. Hence, we get different $R_c$ for different $h$ as shown in Fig. 13(a). From Fig. 13(a), we can determine $h$ from $R_c$. However, there are two possible $h$ for a particular $R_c$: < 450 and >450 nm. For avoiding this ambiguity, the relative dominance of the $c$ parameters with $h$ is considered. Figure 13 shows that $c_{ys} < c_{xp}$ for $h < 400$ nm. However, $c_{ys} = c_{xp}$ for $h \approx 400$ nm. For $h > 450$ nm, $c_{ys} > c_{xp}$ and $c_{xs} > 182.3$ nm. Therefore, if we get two $h$ for the same ratio value, we check the relation between $c_{ys}$ and $c_{xp}$. If $c_{ys} < c_{xp}$, $h < 400$ nm; if $c_{ys} = c_{xp}$, $h \approx 400$ nm; if $c_{ys} > c_{xp}$ and $c_{xs} < 182.3$ nm, $400 < h < 450$ nm; and if $c_{ys} > c_{xp}$ and $c_{xs} > 182.3$ nm, $h > 450$ nm.

**Table 5** $R_c$ values considering different cell shapes and sizes.

| $(w_x, w_y, h)$ (nm) | $R_c$ | $(w_x, w_y, h)$ (nm) | $R_c$ | $(w_x, w_y, h)$ (nm) | $R_c$ |
|---|---|---|---|---|---|
| (1400, 1800, 150) | 1.25 | (1400, 1800, 200) | 1.29 | (1400, 1800, 300) | 1.45 |
| (1600, 1600, 150) | 1.22 | (1800, 1800, 200) | 1.30 | (2000, 1900, 300) | 1.46 |
| (1800, 1800, 150) | 1.26 | (2000, 2000, 200) | 1.30 | (2000, 2000, 300) | 1.46 |
| (1400, 1800, 400) | 1.50 | (1500, 2000, 500) | 1.55 | (1400, 1800, 600) | 1.40 |
| (2000, 1800, 400) | 1.56 | (1800, 1800, 500) | 1.57 | (1600, 1200, 600) | 1.42 |
| (2000, 2000, 400) | 1.53 | (2000, 2000, 500) | 1.50 | (2000, 2000, 600) | 1.44 |

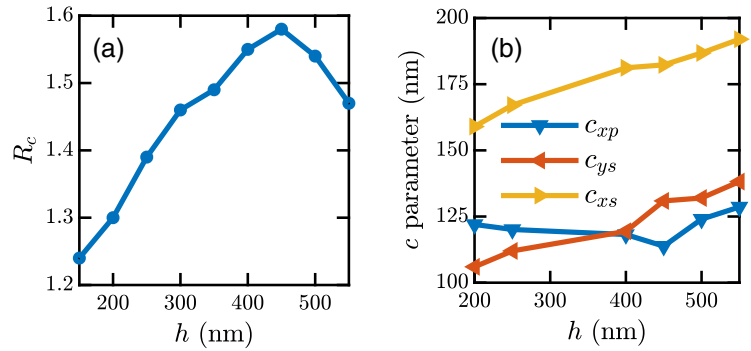

**Fig. 13** Variations of (a) $R_c$ and (b) $c$ parameter against $h$.

### 4.2.2 *Determining fluorophores at random positions*

3D imaging requires determining all three $(x, y, z)$ coordinates of the dipoles located at any position on the cell membrane. We have already discussed about determining $(x, y)$ coordinates and $h$. Now, finding fluorophores located at arbitrary positions on the cell membrane is similar to determining $h$, except that a fluorophore's $R_c$ is different from that at the center at the same fluorophore height. Therefore, two different ratio curves are used: one for the fluorophores at the top ($R_c$) and the other for the fluorophores at random positions on the cell surface ($R'_c$). Table 6 shows $R'_c$ for different cell sizes with fluorophores at arbitrary positions. To determine the $z$ coordinate, $c_{xs}/c_{xp}$ ratio is used as if $w_x > w_y$, but $c_{xs}/c_{yp}$ is used as if $w_x < w_y$.

Table 6 shows $R'_c$ values for a particular $z$ position of fluorophores. $R'_c$ does not vary much with $z$ position irrespective of cell size and fluorophore position on the $xy$ plane. However, the value varies with $z$ positions. Several arbitrary cell samples are simulated with fluorophores at random positions. In each case, $R'_c$ is calculated. Then we took the average of these $R'_c$ values for a particular $z$ coordinate. In this way, we get different values for different $z$ coordinates. Figure 14(a) shows the variation of $R'_c$ with $z$ coordinates of the fluorophores. From this curve, the $z$ coordinates of the fluorophores are determined.

**Table 6** $R'_c$ values considering fluorophores at random positions on the cell surface.

| $(w_x, w_y, h_z)$ (nm) | Fluorophore $(x, y, z)$ (nm) | $R'_c$ |
|---|---|---|
| (1200, 1600, 650) | (460, 0, 400) | 1.46 |
| (1300, 1300, 700) | (350, 350, 400) | 1.47 |
| (1800, 1300, 700) | (420, 350, 400) | 1.47 |
| (1400, 1600, 600) | (400, 200, 500) | 1.32 |
| (1600, 1600, 900) | (500, 300, 500) | 1.38 |
| (1800, 1300, 700) | (420, 350, 500) | 1.33 |

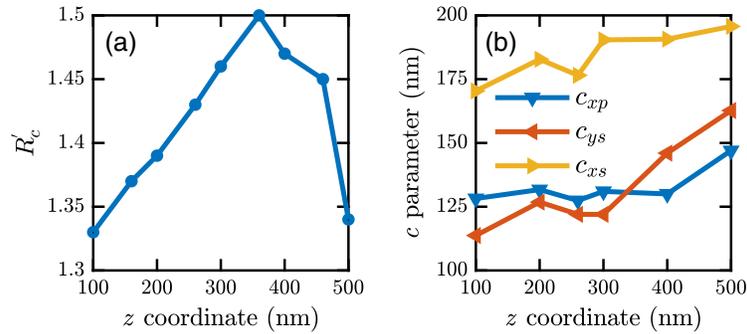

**Fig. 14** Variations of (a) $R'_c$ and (b) $c$ parameters against $z$ coordinates of fluorophores.

Figure 14(a) shows that there are two possible $z$ values for a particular $R'_c$. To avoid this drawback, we use the relations between $c$ parameters. Figure 14(b) shows that $c_{ys} < c_{xp}$ for $z < 350$ nm. For $z > 350$ nm, $c_{ys} > c_{xp}$. Therefore, if we get two $z$ coordinates for the same $R'_c$, the relation between $c_{ys}$ and $c_{xp}$ is used. If $c_{ys} < c_{xp}$, $z < 350$ nm, and if $c_{ys} > c_{xp}$, $z > 350$ nm. For fluorophores with the highest $(x, y)$, which are at the edge of the cell base, $z = 0$.

## 5 3D Imaging of a Cell Using the Proposed Methodology

Now, we show a complete 3D imaging of a cell with $w_x = w_y = 1000$ nm and $h = 400$ nm. The cell is labeled with fluorophores. Using the methods discussed in Sec. 4.1, we determine the fluorophore positions on the $xy$ plane. The edge coordinates are determined, and the base shape is recreated similar to that shown in Fig. 7. We find three distinct SPCE rings in the far-field, i.e., $N_r = 3$. Therefore, $h > 250$ nm. Also the middle ring in the far-field is the strongest. Therefore, according to Table 2, 250 nm $\leq h \leq 800$ nm. Furthermore, the middle ring occurs at ∼35 deg. Therefore, 350 nm $\leq h \leq 700$ nm. The estimated $h$ is narrowed down to 350 to 700 nm combining all three possibilities. To determine $h$ more precisely, we follow the procedures discussed in the latter part of Sec. 4.2.1. The $c_{xs}$, $c_{xp}$ parameters obtained from Gaussian fit on the intensity distributions are 181.2 and 118.2 nm. Therefore, $R_c = 1.53$. Using the $R_c$ value in Fig. 13, we find $h = 385.2$ nm, with a 3.7% error from the original $h = 400$ nm. Then the $z$ coordinates of other fluorophores are determined using the procedures discussed in Sec. 4.2.2. Table 7 and Fig. 15 present the fluorophore coordinates $(x, y, z)$ and the recreated 3D cell shape. Only a few coordinates are shown in Table 7 as examples. It is noted that the RMSE for $(x, y, z)$ coordinates are 2.32%, 3.63%, and 6.14%, respectively.

**Table 7** Comparison of the actual cell coordinates and the recreated coordinates using the proposed methodology. Here only few of the points in the first quadrant are shown as the simulated structure is symmetrical in the $x$ and $y$ directions.

| Actual $(x, y, z)$ (nm) | Calculated $(x, y, z)$ (nm) | Error (%) | | |
|---|---|---|---|---|
| | | $x$ | $y$ | $z$ |
| (1000, 0, 0) | (983.5, 0, 0) | 1.65 | 0 | 0 |
| (755, 0, 260) | (783.2, 0, 280.3) | 3.74 | 0 | 7.8 |
| (433.65, 0, 360) | (445, 0, 353) | 2.62 | 0 | 1.94 |
| (0, 0, 400) | (0, 0, 385.2) | 0 | 0 | 3.7 |
| (650, 650, 160) | (640.6, 680.7, 178) | 1.45 | 4.72 | 11.25 |
| (300, 300, 360) | (315.3, 320, 330) | 5.1 | 6.67 | 8.33 |

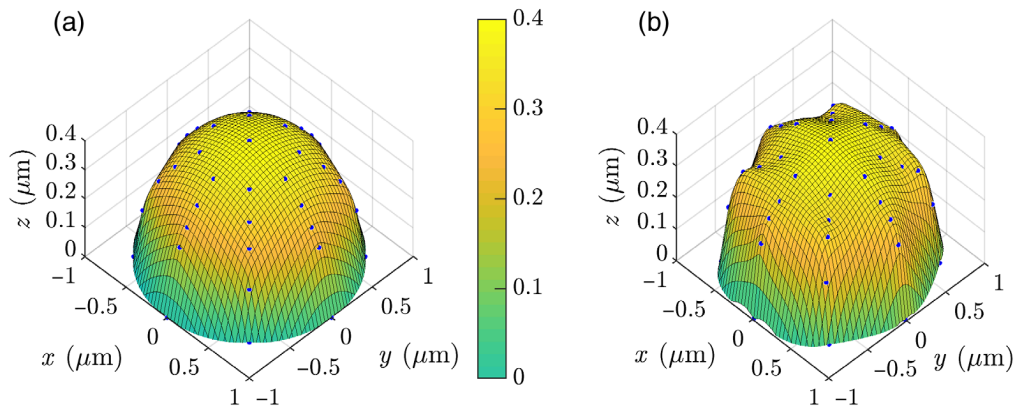

**Fig. 15** (a) Cell shape and (b) 3D image of the cell shape using the proposed technique.

## 6 Conclusion

This work shows that SPCE features can be recorded and postprocessed to create 3D images of biological cells. Considering the cell's variable size and shape, the RMSE of the created image remains within a few percentages, which compares well with expensive state-of-the-art techniques. The results are remarkable considering the simplicity of the proposed approach. The images will be enhanced if the fluorophore count increases. An increasing number of fluorophores will enable us to label more points of the cell surface, helping resolve much smaller structural details. The primary motive of our work was to develop a cheap and compact cell imaging technique. Conventional methods for cell imaging are complicated and expensive. This technique, if produced, will be available at a fractional cost in a hand-held device.

## Disclosures

The authors declare no conflicts of interest.

## Code, Data, and Materials Availability

The proposed methodology is simulated using FDTD simulation technique in Lumerical FDTD Solutions (Release: 2018a, version: 8.19.1584). The simulated data have been analyzed using codes written in MATLAB. The authors will provide additional details upon request.

## References

1. E. Paluch and C.-P. Heisenberg, "Biology and physics of cell shape changes in development," *Curr. Biol.* **19**(17), R790–R799 (2009).
2. Y. Xiong et al., "Automated characterization of cell shape changes during amoeboid motility by skeletonization," *BMC Syst. Biol.* **4**, 33 (2010).
3. "Animal cells and their shapes—Science Learning Hub," https://www.sciencelearn.org.nz/resources/498-animal-cells-and-their-shapes (2012).
4. L. Eldridge, "Cancer cells vs. normal cells: how are they different," https://www.verywellhealth.com/cancer-cells-vs-normal-cells-2248794 (2022).
5. A. Jemal et al., "Global cancer statistics," *CA Cancer J. Clin.* **61**(2), 69–90 (2011).
6. B. Mondeja et al., "Sars-cov-2: preliminary study of infected human nasopharyngeal tissue by high resolution microscopy," *Virol. J.* **18**(1), 1–8 (2021).
7. C. G. Golding et al., "The scanning electron microscope in microbiology and diagnosis of infectious disease," *Sci. Rep.* **6**, 26516 (2016).
8. S.-Y. Tang et al., "High resolution scanning electron microscopy of cells using dielectrophoresis," *PLoS One* **9**, e104109 (2014).

9. B. Titze and C. Genoud, "Volume scanning electron microscopy for imaging biological ultrastructure," *Biol. Cell* **108**(11), 307–323 (2016).

10. X. Zhang et al., "Label-free live-cell imaging of nucleic acids using stimulated Raman scattering microscopy," *ChemPhysChem* **13**(4), 1054–1059 (2012).

11. W. Liu et al., "Raman microspectroscopy of nucleus and cytoplasm for human colon cancer diagnosis," *Biosens. Bioelectron.* **97**, 70–74 (2017).

12. R. Smith, K. L. Wright, and L. Ashton, "Raman spectroscopy: an evolving technique for live cell studies," *Analyst* **141**(12), 3590–3600 (2016).

13. D. Axelrod, "Total internal reflection fluorescence microscopy in cell biology," *Traffic* **2**(11), 764–774 (2001).

14. B. A. Millis, "Evanescent-wave field imaging: an introduction to total internal reflection fluorescence microscopy," *Methods Mol. Biol.* **823**, 295–309 (2012).

15. R. F. Thompson et al., "An introduction to sample preparation and imaging by cryo-electron microscopy for structural biology," *Methods* **100**, 3–15 (2016).

16. K. Eberhardt et al., "Advantages and limitations of Raman spectroscopy for molecular diagnostics: an update," *Expert Rev. Mol. Diagn.* **15**(6), 773–787 (2015).

17. R. Roy, S. Hohng, and T. Ha, "A practical guide to single-molecule FRET," *Nat. Methods* **5**(6), 507–516 (2008).

18. F. Stefani et al., "Surface-plasmon-mediated single-molecule fluorescence through a thin metallic film," *Phys. Rev. Lett.* **94**(2), 023005 (2005).

19. J. Borejdo et al., "Application of surface plasmon coupled emission to study of muscle," *Biophys. J.* **91**(7), 2626–2635 (2006).

20. Q. Liu et al., "Surface plasmon coupled emission in micrometer-scale cells: a leap from interface to bulk targets," *J. Phys. Chem. B* **119**(7), 2921–2927 (2015).

21. J. R. Lakowicz et al., "Directional surface plasmon-coupled emission: a new method for high sensitivity detection," *Biochem. Biophys. Res. Commun.* **307**(3), 435–439 (2003).

22. J. R. Lakowicz, "Radiative decay engineering 3. Surface plasmon-coupled directional emission," *Anal. Biochem.* **324**(2), 153–169 (2004).

23. I. Gryczynski et al., "Effects of sample thickness on the optical properties of surface plasmon-coupled emission," *J. Phys. Chem. B* **108**(32), 12073–12083 (2004).

24. I. Gryczynski et al., "Radiative decay engineering 4. Experimental studies of surface plasmon-coupled directional emission," *Anal. Biochem.* **324**(2), 170–182 (2004).

25. N. Calander, "Surface plasmon-coupled emission and Fabry–Pérot resonance in the sample layer: a theoretical approach," *J. Phys. Chem. B* **109**(29), 13957–13963 (2005).

26. S.-H. Cao et al., "Surface plasmon-coupled emission: what can directional fluorescence bring to the analytical sciences?" *Annu. Rev. Anal. Chem.* **5**, 317–336 (2012).

27. H. Ma et al., "Fast and precise 3D fluorophore localization based on gradient fitting," *Sci. Rep.* **5**(1), 14335 (2015).

28. A. Small and S. Stahlheber, "Fluorophore localization algorithms for super-resolution microscopy," *Nat. Methods* **11**(3), 267–279 (2014).

29. S. Z. Uddin and M. A. Talukder, "Imaging of cell membrane topography using Tamm plasmon coupled emission," *Biomed. Phys. Eng. Express* **3**(6), 065005 (2017).

30. N. Calander, "Theory and simulation of surface plasmon-coupled directional emission from fluorophores at planar structures," *Anal. Chem.* **76**(8), 2168–2173 (2004).

31. J. Enderlein and T. Ruckstuhl, "The efficiency of surface-plasmon coupled emission for sensitive fluorescence detection," *Opt. Express* **13**, 8855–8865 (2005).

32. N. H. T. Tran et al., "Reproducible enhancement of fluorescence by bimetal mediated surface plasmon coupled emission for highly sensitive quantitative diagnosis of double-stranded DNA," *Small* **14**(32), 1801385 (2018).

33. C. Heck, "Gold and silver nanolenses self-assembled by DNA origami," PhD Thesis, Universität Potsdam (2017).

34. K. Aslan and C. D. Geddes, "Directional surface plasmon coupled luminescence for analytical sensing applications: which metal, what wavelength, what observation angle?" *Anal. Chem.* **81**(16), 6913–6922 (2009).

35. J. R. Lakowicz, *Principles of Fluorescence Spectroscopy*, Springer (2006).

36. A. S. Kristoffersen et al., "Testing fluorescence lifetime standards using two-photon excitation and time-domain instrumentation: rhodamine b, coumarin 6 and lucifer yellow," *J. Fluoresc.* **24**(4), 1015–1024 (2014).

37. M. A. Talukder, C. R. Menyuk, and Y. Kostov, "Distinguishing between whole cells and cell debris using surface plasmon coupled emission," *Biomed. Opt. Express* **9**(4), 1977–1991 (2018).

38. J. B. Grimm et al., "A general method to fine-tune fluorophores for live-cell and in vivo imaging," *Nat. Methods* **14**(10), 987–994 (2017).

39. R. W. Sabnis, *Handbook of Biological Dyes and Stains: Synthesis and Industrial Applications*, John Wiley & Sons, Hoboken, New Jersey (2010).

40. M. Snare et al., "The photophysics of rhodamine b," *J. Photochem.* **18**(4), 335–346 (1982).

41. T. C. Brelje, M. W. Wessendorf, and R. L. Sorenson, "Multicolor laser scanning confocal immunofluorescence microscopy: practical application and limitations," *Cell Biol. Appl. Confocal Microsc.* **70**, 165–244 (2002).

42. H. Edelhoch, S.-Y. Cheng, and G. Irace, "Spectroscopic properties of rhodamine b-labeled thyroid hormone," *Ann. N.Y. Acad. Sci.* **366**(1), 253–264 (1981).

43. E. D. Palik, *Handbook of Optical Constants of Solids*, Academic Press, Burlington (1997).

44. Y. Sung et al., "Optical diffraction tomography for high resolution live cell imaging," *Opt. Express* **17**, 266–277 (2009).

45. J. Chang et al., "Automated tuberculosis diagnosis using fluorescence images from a mobile microscope," *Lect. Notes Comput. Sci*, **7512**, 345–352 (2012).

46. W. Choi et al., "Tomographic phase microscopy," *Nat. Methods* **4**(9), 717–719 (2007).

47. P. Y. Liu et al., "Cell refractive index for cell biology and disease diagnosis: past, present and future," *Lab Chip* **16**(4), 634–644 (2016).

48. T. Reimer, M. Solis-Nepote, and S. Pistorius, "The impact of the inverse chirp z-transform on breast microwave radar image reconstruction," *Int. J. Microwave Wireless Technol.* **12**(9), 848–854 (2020).

49. M. Chen, S. H. Cao, and Y. Q. Li, "Surface plasmon-coupled emission imaging for biological applications," *Anal. Bioanal. Chem.* **412**, 6085–6100 (2020).

50. G. de Vito et al., "Effects of excitation light polarization on fluorescence emission in two-photon light-sheet microscopy," *Biomed. Opt. Express* **11**, 4651–4665 (2020).

51. M. Wagner et al., "Fluorescence and polarization imaging of membrane dynamics in living cells," *Proc SPIE* **7176**, 717607 (2009).

52. S. Brasselet, "Polarization-resolved nonlinear microscopy: application to structural molecular and biological imaging," *Adv. Opt. Photonics* **3**, 205 (2011).

53. S. Z. Uddin, M. R. Tanvir, and M. A. Talukder, "A proposal and a theoretical analysis of an enhanced surface plasmon coupled emission structure for single molecule detection," *J. Appl. Phys.* **119**(20), 204701 (2016).

54. C. D. Geddes et al., "Directional surface plasmon coupled emission," *J. Fluoresc.* **14**(1), 119–123 (2004).

**Anik Mazumder** received his bachelor's of science degree in electrical and electronic engineering (EEE) from Bangladesh University of Engineering and Technology (BUET) in 2021. After graduation, he worked as an adjunct lecturer at BUET for nine months. Currently, he is working as a lecturer in the Department of Computer Science and Engineering (CSE) at the United International University (UIU), Bangladesh. He is pursuing a master's degree in nanophotonics.

**Mohammad Mozammal** received his bachelor's of science degree in electrical and electronic engineering (EEE) from Bangladesh University of Engineering and Technology (BUET) in 2021. Since graduation, he has been working as an adjunct lecturer in the Department of Electrical Electronic and Communication Engineering (EECE) at the Military Institute of Science and Technology, Bangladesh. His research interests are nanophotonics, biosensing, plasmonics, and solid-state electronics.

**Muhammad Anisuzzaman Talukder** joined the Electrical and Electronic Engineering Department of the Bangladesh University of Engineering and Technology (BUET) as a lecturer in early 2001 and has been serving as a professor since 2014. From 2016 to 2018, he was a distinguished Marie-Curie Individual Fellow at University of Leeds, United Kingdom. From 2013 to 2015, he was an honorary fellow at the Hong Kong Polytechnic University and a visiting academic fellow at the City University London, United Kingdom, in 2013.